\documentclass[12pt]{emulateapj}
\shorttitle{Binary BD+20 307}
\shortauthors{Weinberger}
\submitted{Accepted to ApJL}

\begin{document}
\title{On the Binary Nature of Dust-encircled BD+20 307}
\author{A. J. Weinberger}
\affil {Department of Terrestrial Magnetism, Carnegie Institution of
Washington\\
5241 Broad Branch Road NW, Washington, DC 20015}
\email{weinberger@dtm.ciw.edu}

\begin{abstract}
Three epochs of high resolution spectra of the star \object{BD+20 307}
show that it is a short period ($\sim$3.4 day) spectroscopic binary of
two nearly identical stars. Surprisingly, the two stars, though
differing in effective temperature by only $\sim$250~K and having a mass
ratio of 0.91, show very different Li line equivalent widths. A Li 6707\AA\ line
is only detected from the primary star, and it is weak. This star is
therefore likely to be older than 1 Gyr. If so, the large amount of hot
circumbinary dust must be from a very large and recent, but very late
evolutionarily, collision of planetesimals.
\end{abstract}


\section{Introduction}

Debris disks older than $\sim$10 Myr containing dust at temperatures
$>$100~K are extremely rare \citep{Aumann1991,Silverstone06}. When warm
dust does appear, it is likely to be from a stochastic event, perhaps
akin to our own solar system's ``Late Heavy Bombardment,'' at about 600
Myr after formation. Late episodes of dust production may signal the
presence of a planetary system undergoing architectural reconfiguration
\citep{Gomes2005}.

BD+20 307 is one of the few examples of a non-young star with
hot debris \citep{Song2005}. It has a ring of dust at $\sim$0.5 AU
\citep{Weinberger08}.  In order to understand the implications of
the large amount of close-in dust, it would help greatly to know the age
of the star.  \citet{Song2005} used the Li 6707\AA\
equivalent width and chromospheric activity to suggest an age
of $\sim$300 Myr.  

New observations reported here show that BD +20 307 is actually a
spectroscopic binary. I reexamine the evidence for the age of the star.

\section{Observations}

The Magellan Inamori Kyocera Echelle (MIKE) spectrograph on the Clay
(Magellan II) telescope was used to observe BD +20 307 on three
consecutive nights -- 2007 October 24-26 (UT). The 0.35'' wide $\times$
5'' long slit provided a resolution of about 55,000 at wavelengths 3400
-- 7250 \AA.  Seeing was $\sim$0.5'' on the first two nights and
0.8--1.2'' on the third. On all three nights, data were obtained with an
iodine cell in place to facilitate looking for planets around the star,
and on the first night an observation without the iodine cell was also
obtained. I do not use the iodine lines for the radial velocity
analyses that follow.  An observing log is given in Table
\ref{tab_observlog}.

\begin{deluxetable}{llc}
\tablewidth{0pt}
\tablecaption{Observing Log \label{tab_observlog}}
\tablehead{
\colhead{Date (UT)} &\colhead{Time}  &\colhead{Int. Time (s)}}
\startdata
2004 Aug 24         &08:28:20        &700\\
2007 Oct 24         &04:35:20        &600\\
2007 Oct 25         &04:45:43        &600\\
2007 Oct 26         &04:12:33        &600\\
\enddata
\end{deluxetable}

The spectra were flattened, extracted and wavelength calibrated using
the MIKE pipeline written by D. Kelson with methods described in
\citet{Kelson2000,Kelson2003,Kelson2006}. The two observations from the
first night were averaged. The signal-to-noise ratio (S/N) per pixel was
$>$100 for wavelengths $>$4000 \AA\ on the first two nights, except in
the region of maximal iodine absorption around 5000\AA. The S/N was
about 50\% worse on the third night due to the worse seeing.

BD +20 307 was also observed on 2004 August 24 with the echelle
spectrograph on the 2.5 m du~Pont Telescope at Las Campanas
Observatory. These data cover wavelengths $\sim$4000--9000\AA\ and have a
resolution of about 25,000 and S/N of 30--100. The data were extracted
and calibrated using standard IRAF tasks.

Heliocentric and barycentric velocities were calculated with the RVSAO
package in IRAF.

\section{Results \label{results}}

\begin{figure}
\plotone{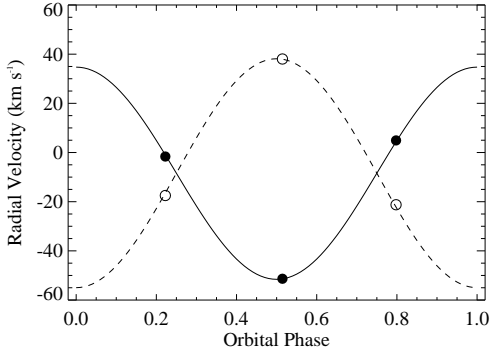}
\caption{Velocities for the primary (filled circles) and secondary (open
circles) components of the binary phased with an orbital period of
3.448 d. The residuals between the fit orbit and the data are $<$1 km
s$^{-1}$ on each point. \label{fig_binaryvel}}
\end{figure}

Two sets of lines are clearly visible in all three nights of MIKE
data. To obtain the velocities of the double-lined spectroscopic binary,
cross-correlations with a synthetic spectrum with effective temperature
6000 K and log(g)=5.0 were performed. This spectrum was generated using
R. O. Gray's SPECTRUM code and line list\footnotemark[1] and a Castelli-Kurucz model
atmosphere with solar metalicity\footnotemark[2]. 
The xcsao package in IRAF was used to compute the cross-correlations, and
the two peaks were fit with parabolas in IDL. The uncertainty in the
velocities was computed as the standard deviation of the velocities in
the 40 different orders used.  Results are reported in
Table \ref{tab_binaryvel} and shown in Figure \ref{fig_binaryvel}. On
all three nights, the primary star produced a higher cross correlation
peak.

\footnotetext[1]{\url{http://www.phys.appstate.edu/spectrum/detail.html}}

\footnotetext[2]{http://kurucz.harvard.edu/grids/gridP00aODFNEW/ap00k0odfnew.dat}

\begin{deluxetable}{llll}
\tablewidth{0pt}
\tablecaption{Measured radial velocities \label{tab_binaryvel}}
\tablehead{
\colhead{} &\multicolumn{3}{c}{HJD}\\
\colhead{Component} &\colhead{2454397.6969} &\colhead{2454398.7041}&\colhead{2454399.6810}}
\startdata
Primary   & -1.63 $\pm$ 0.46 & -51.34 $\pm$ 0.62  & 4.92 $\pm$ 0.79\\
Secondary &-17.49 $\pm$ 0.85 & 38.03  $\pm$ 0.64 &-21.24 $\pm$ 0.67\\
\enddata
\end{deluxetable}

The same cross-correlation was done for the lower resolution du~Pont
spectrum. A double-peaked cross-correlation appears only for the lowest
two orders (4000--4150\AA). I do not consider this detection of the
binary reliable, and do not include these RV data in the analyses which follow.

\begin{figure}
\plotone{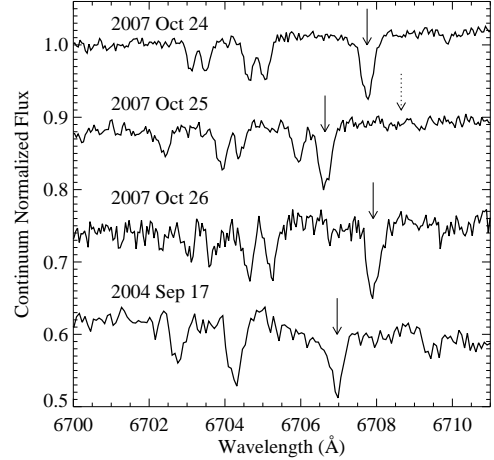}
\caption{A Lithium 6707\AA\ line is detected from the primary star in
all four epochs (solid arrows). These spectra have been continuum normalized
and offset for clarity. The best upper limit (14 m\AA, 3$\sigma$) on the secondary's Li line
can be obtained on 25 Oct; the dashed arrow shows where it should
be located. \label{fig_lithium}}
\end{figure}

On all nights a Li 6707\AA\ line was detected from the primary star
(Figure \ref{fig_lithium}).  The equivalent widths were computed using
direct integration over the lines relative to the combined continua from
the two stars. Uncertainties from the pipeline reduction were used to
give the statistical uncertainty. An additional systematic uncertainty
was estimated by choosing different methods of finding the continuum and
recomputing the equivalent widths.  A 3$\sigma$ upper limit on the
secondary's Li line was placed using the data from 2007 October 25, when
the two stars were separated by 89 km~$\rm s^{-1}$ (2 \AA). These
equivalent widths are given in the first two columns of Table \ref{tab_Li_eqw}.

\begin{deluxetable}{lllll}
\tablewidth{0pt}
\tablecaption{Equivalent width (in m\AA) of Li 6707\AA\ line
\label{tab_Li_eqw}}
\tablehead{
\colhead{}     &\multicolumn{2}{c}{Combined Continuum}&\multicolumn{2}{c}{Separate Continuum}\\
\colhead{Date} &\colhead{Primary} &\colhead{Secondary} &\colhead{Primary} &\colhead{Secondary}}
\startdata
2004 Aug 24 &41 $\pm$ 3  &\nodata          &70 $\pm$ 5 &\nodata \\
2007 Oct 24 &35 $\pm$ 2  &\nodata          &60 $\pm$ 4 &\nodata \\
2007 Oct 25 &34 $\pm$ 2  &$<$6 (3$\sigma$) &58 $\pm$ 4 &$<$14 (3$\sigma$) \\
2007 Oct 26 &33 $\pm$ 2  &\nodata          &56 $\pm$ 4 &\nodata \\
\enddata
\end{deluxetable}

The continuum normalized spectrum on 2007 October 25, which has the
maximum separation of the two stars, was fit with a combination of
synthetic spectra calculated with SPECTRUM from Castelli-Kurucz model
atmospheres with solar metalicity. Free parameters were two effective
temperatures, a single $v {\rm sin} i$, a single log(g), and two
normalizations. One blue and one red region of the spectrum were fit --
4099--4360 \AA\ and 6282-6549 \AA.  The best fit in both cases had
T$_{\rm eff}$=6500 K and 6250~K for the primary and secondary stars,
respectively and log(g)=5.0. Contours of chi-square indicate the
uncertainty is within 250 K (the gridding of the models) in T$_{\rm
  eff}$.  The lines are measurably broader than the ThAr calibration
lamp lines at the same wavelengths. The best fit models had $v {\rm sin}
i = 10 \pm 1$ km s$^{-1}$.

To compute the stars' Li equivalent widths relative to their own stellar
continua, the flux ratio of the two stars must be obtained at 6707\AA.
Synthetic spectra were fit as above to the region at 6645 --
6835\AA. The best fit flux ratio was 1.397 $\pm$ 0.007. However, this
statistical uncertainty probably underestimates the systematics from how
the models are calculated.  Applying this flux ratio to the measured
equivalent widths makes the primary and secondary equivalent widths
increase by factors of 1.709 and 2.387, respectively. The computed equivalent
widths are given in the last two columns of Table \ref{tab_Li_eqw}.

\begin{figure}
\plotone{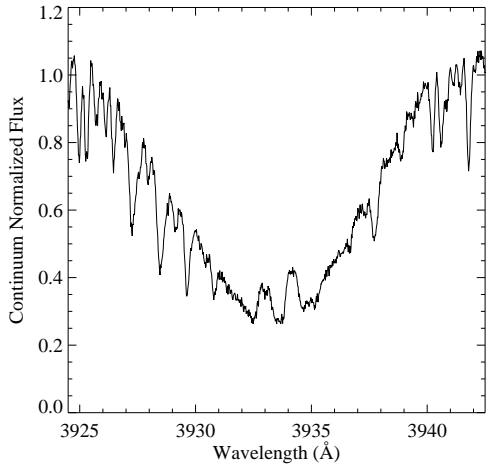}
\caption{Closeup of the calcium K line from 2007 October 25 showing a weak central reversal
for each star. The secondary star's reversal (red-ward line) is
stronger. The spectrum was continuum normalized.\label{fig_calcium}}.
\end{figure}

The spectra were examined for evidence of chromospheric activity.  Both
stars show weak central reversals on their Ca H and K lines (Figure
\ref{fig_calcium}) that change in velocity along with the stars. The Balmer lines show no
central reversals.

The velocities were fit for the stellar mass ratio (secondary/primary)
yielding 0.91 $\pm$ 0.02 and the line of sight radial velocity yielding
$\rm \gamma = -8.8 \pm 0.6\ km\ s^{-1}$.

Although the small number of observations prohibit the calculation of
the full binary orbit, I tested whether a circular orbit could fit the
velocities. The best fit produced reasonable residuals ($<$1 km~$\rm
s^{-1}$) for an orbit with a period of 3.448 days, $\gamma$=-8.4 km~$\rm
s^{-1}$, consistent with the fit above, and mass ratio 0.93, again
consistent with the fit above. This orbit, shown in Figure
\ref{fig_binaryvel} yields $m {\rm sin^3}i$ for each star -- 0.13 and
0.12 for the primary and secondary stars respectively. I estimate the
true masses of the stars as 1.3 and 1.2 M$_\odot$ based on the effective
temperatures (6500 and 6250 K) obtained in the spectral fit described in
\S~3 and the MK spectral-type calibration given in
\citet{DrillingLandolt}; these masses are uncertain by $\sim$0.1
M$_\odot$. Calculating the inclination from these masses yield 27.8 $\pm$ 0.8$^\circ$.

The fit, assuming the inclination of 28$^\circ$, gives semi-major axes
for the binary orbits of 4.4 and 4.7~ $\times$ 10$^6$ km. The stars
should have radii of 1.3--1.6 R$_\odot$, so their separation is about 9 
stellar radii.

\section{Discussion}

Of primary interest for understanding the dusty debris disk around BD+20
307 is the age of the system.

Although the orbit is not well measured, a reasonable fit is obtained with
e=0. Binaries of any age with periods less than $\sim$8 days are found
in circular orbits \citep{Mathieu1992,Melo2001}. So, e=0 is consistent with the 3.4
day period and is no indication of the age of the system.

The formation of close binaries may depend on the presence of another
star in the system; \citet{Tokovinin} find that 80\% of binaries with
periods $<$7 days are actually in higher order multiple systems. It is
unknown whether BD+20 307 has another, wider, component, but this should
be investigated.

Taking $\rm sin i =28^\circ$ from the spectral modeling, the rotational
velocity of the stars is 21 km s$^{-1}$. A standard radius for a
T$_{eff}$=6250~K star would be 1.3 R$_\odot$, but solar-like stars with
active chromospheres have somewhat larger radii than their non-active
counterparts \citep{Torres06}. Stellar radii of 1.3 -- 1.6 R$_\odot$
correspond to rotational periods of 3 -- 3.7~d.  The close agreement
between the rotation period and the orbital period suggests that the
stars are tidally synchronized.

The timescale for synchronization for stars of mass similar to the Sun
is given by \citet{Zahn}, and for equal mass components is $\approx \rm
10^4 P^4 yr$ where P is the orbital period. Given a period of 3.5 d,
t$_{\rm synch}$ is 100 Myr. This provides a lower limit to the age of
the system.

Tidal coupling in a close binary can give rise to enhanced magnetic
activity \citep{Rottler2002}. So, the central reversals observed in
both stars' Ca H and K lines need not arise from youth but could be from
their interaction.

For stars with deep convective zones, there is a tight correlation
between lithium abundance with age and spectral type, but for stars with
T$_{eff}$ $\sim$ 6500~K, there is little Li depletion by the age of the
Hyades \citep{Soderblom90}. A typical Li equivalent width for a star
with B-V of 0.5 in the Hyades is 90 m\AA, 1.5 times that for BD+20
307.  Furthermore, tidally locked systems in the Hyades show inhibited
Li depletion \citep{Barrado1996}.

The situation for older stars is more complex. Stars even hotter than
the Sun apparently do deplete Li after reaching the main sequence.
\citet{Sestito} summarize the Li abundances in open clusters of various
ages, including the wide spread in abundances among members of old ($>$1
Gyr) clusters.  Hot stars can show substantial
depletions. For example in the 1 Gyr old cluster NGC 3960, stars with
$B-V \approx 0.5$ (T$_{eff} \sim$ 6250~K) show a range EW(Li) of $<$10
to 80 m\AA\ \citep{Prisinzano07}.

Initial metalicity can have a large effect on main sequence Li
abundance, but given their tight orbit, the two stars in BD+20 307
presumably formed from the same molecular cloud material at the same
metalicity. \citet{Martin2002} found that $\sim$20\% of wide field
binaries showed disparate Li. Stars may also be polluted by Li and other
elements during planetary evolution \citep[e.g.][]{Gonzalez}, which could
account for differences in Li abundance between wide binaries and within
clusters that should have the same initial metalicity.  Especially for
stars with small convective zones, like the components of BD+20 307,
material deposited should last for the entire main sequence lifetime.
The tightness of the BD+20 307 orbit means that any planets must be
circumbinary. If disparate pollution were to occur, it might be expected
to favor the more massive star. It is indeed the primary star that shows
the greater Li line strength.

Taken together, the modest Li line strength of the primary and very low
Li line strength for the secondary argue for an old age of the system,
certainly greater than that of the Hyades and probably $>$ 1 Gyr. In
this case, the dust generating event happened quite late.  In the Solar
System, the last era of major planetesimal removal happened at an age
600-800 Myr \citep{Strom05}.

It is further tempting to speculate on what happens to the dust ring
over time. Poynting-Robertson and solar wind drag should cause the
grains at 0.55 AU to fall onto the central stars on timescales $<$1000
yr \citep{Weinberger08}. Meteorites in the Solar System contain Li, no
matter where they formed in the original proto-planetary disk
\citep{Seitz07}. Thus the opportunity seems to exist for substantial
stellar pollution with Li by the dust around BD+20 307.  Perhaps this
dust is being swept up by a body orbiting between the ring and the
binary and thus cannot pollute the surface of the stars. A detailed
abundance study of the stars could reveal the extent of any pollution.

\section{Conclusions}

BD+20 307 is a spectroscopic binary composed of two similar stars of
late F spectral-type in a short, rotationally synchronized orbit of
$\sim$3.4 days.  Standard chromospheric-age calibrations for single
solar-type stars are therefore not applicable to this system. The binary
is likely quite old, $>$ 1 Gyr, in order to allow the two stars to
develop quite different lithium equivalent widths.  The collision that produced
the circumbinary dust ring must have happened recently, given the short
timescale for removal of material at 0.5 AU, and could be expected to
pollute the envelopes of the stars. That they have such low lithium
could be evidence that the dust does not actually reach them.

\acknowledgements 

I acknowledge Paul Butler and Mercedes Lopez-Morales for assistance in
setting up MIKE, John Debes for performing the observations, Johanna
Teske for running the data through the pipeline, and George
Preston for obtaining the du Pont spectra. Thanks also to Lopez-Morales
and Evgenya Shkolnik for helpful discussions and Ben Zuckerman for
comments on the manuscript. This work was supported in
part by a Spitzer Space Telescope grant through JPL and the Carnegie
node of the NASA Astrobiology Institute.

{\it Facilities:} \facility{Magellan:Clay (MIKE)}


\begin{thebibliography}{1}

\bibitem[Aumann \& Probst(1991)]{Aumann1991} Aumann, H.~H.,
\& Probst, R.~G.\ 1991, \apj, 368, 264

\bibitem[Barrado y Navascues \& Stauffer(1996)]{Barrado1996}
Barrado y Navascues, D., \& Stauffer, J.~R.\ 1996, \aap, 310, 879

\bibitem[Drilling \& Landolt(2000)]{DrillingLandolt} Drilling, J. S. \&
  Landolt, A. U. 2000, in Allen's Astrophysical Quantities,
  ed. A. N. Cox (New York: AIP Press), 381

\bibitem[Gomes et al.(2005)]{Gomes2005} Gomes, R., Levison,
H.~F., Tsiganis, K., \& Morbidelli, A.\ 2005, \nat, 435, 466

\bibitem[Gonzalez(2006)]{Gonzalez} Gonzalez, G. 2006, \pasp, 118, 1494

\bibitem[Kelson et al.(2000)]{Kelson2000} Kelson, D.~D.,
Illingworth, G.~D., van Dokkum, P.~G., \& Franx, M.\ 2000, \apj, 531,
159
\bibitem[Kelson(2003)]{Kelson2003} Kelson, D.~D.\ 2003, \pasp,
115, 688

\bibitem[Kelson et al.(2006)]{Kelson2006} Kelson, D.~D.,
Illingworth, G.~D., Franx, M., \& van Dokkum, P.~G.\ 2006, \apj, 653,
159

\bibitem[Mart{\'{\i}}n et al.(2002)]{Martin2002} Mart{\'{\i}}n,
E.~L., Basri, G., Pavlenko, Y., \& Lyubchik, Y.\ 2002, \apj, 579, 437

\bibitem[Mathieu et al.(1992)]{Mathieu1992} Mathieu, R.~D., Latham, 
D.~W., Mazeh, T., Duquennoy, A., Mayor, M., 
\& Mermilliod, J.-C.\ 1992, in Binaries as
Tracers of Stellar Formation. Editors, A. Duquennoy \& M. Mayor
(Cambridge: Cambridge University Press), 278 

\bibitem[Melo et al.(2001)]{Melo2001} Melo, C. H. F., Covino, E.,
  Alcala, J. M. \& Torres, G. 2001, \aap, 378, 898

\bibitem[Prisinzano \& Randich(2007)]{Prisinzano07} Prisinzano, L. \&
  Randich, S. 2007, \aap, 475, 539

\bibitem[Rottler et al.(2002)]{Rottler2002} Rottler, L., Batalha, C., Young, A. \& Vogt,
  S. 2002, \aap, 392, 535

\bibitem[Seitz et al.(2007)]{Seitz07} Seitz, H.-M., Brey,
G.~P., Zipfel, J., Ott, U., Weyer, S., Durali, S., \& Weinbruch, S.\
2007, Earth and Planetary Science Letters, 260, 582

\bibitem[Sestito \& Randich(2005)]{Sestito} Sestito, P. \& Randich,
  S. 2005,\aap, 442, 615

\bibitem[Silverstone et al.(2006)]{Silverstone06} Silverstone, M.D. et
  al. 2006, \apj, 639, 1138

\bibitem[Soderblom et al.(1990)]{Soderblom90} Soderblom, D.~R.,
Oey, M.~S., Johnson, D.~R.~H., \& Stone, R.~P.~S.\ 1990, \aj, 99, 595


\bibitem[Song et al.(2005)]{Song2005} Song, I., Zuckerman, B.,
Weinberger, A. J. \& Becklin, E. E. 2005, \nat, 436, 363

\bibitem[Strom et al.(2005)]{Strom05}Strom, R. G., Malhotra, R., Ito,
  T., Yoshida, F., \& Kring, D. A. 2005, Science, 309, 1847

\bibitem[Tokovinin et al.(2006)]{Tokovinin} Tokovinin, A.  Thomas, S.,
  Sterzik, M. \& Udry, S. 2006, \aap, 450, 681

\bibitem[Torres et al.(2006)]{Torres06} Torres, G., Lacy,
  C.~H., Marschall, L.~A., Sheets, H.~A., \& Mader, J.~A.\ 2006, \apj,
  640, 1018

\bibitem[Weinberger et al.(2008)]{Weinberger08} Weinberger, A. J.,
Becklin, E. E., Song, I. \& Zuckerman, B. 2008, \apj, submitted


\bibitem[Zahn(1992)]{Zahn} Zahn, J.~P.\ 1992, in Binaries as
Tracers of Stellar Formation. Editors, A. Duquennoy \& M. Mayor
(Cambridge: Cambridge University Press), 253

\end{thebibliography}
\end{document}